\documentclass[aip,jcp,reprint,longbibliography%
               ]{revtex4-1}

\usepackage{graphicx}
\usepackage{xcolor}
\usepackage{amsmath}

\begin{document}


\title{Classifying destructive quantum interference in molecular junctions: Towards molecular quantum rulers}

\author{Lukas Krieger}
\affiliation{Institute of Physics, University of Augsburg, 86135 Augsburg, Germany}

\author{Gert-Ludwig Ingold}
\affiliation{Institute of Physics, University of Augsburg, 86135 Augsburg, Germany}

\author{Fabian Pauly}
\email{fabian.pauly@uni-a.de}
\affiliation{Institute of Physics, University of Augsburg, 86135 Augsburg, Germany}
\affiliation{Center for Advanced Analytics and Predictive Sciences,
University of Augsburg, 86135 Augsburg, Germany}


\date{\today}

\begin{abstract}
Destructive quantum interference in molecular junctions might be used to build molecular quantum rulers, allowing to quantify changes in external control parameters electrically. For this reason, it is important to understand which patterns of destructive quantum interference can occur inside the electronic excitation gap of a molecule, coupled to conducting electrodes. By considering a four-level model, we show that much more complex destructive quantum interference behavior can arise than expected for just two levels.
We classify the destructive quantum interferences analytically and show that they may even occur in regions forbidden by the standard orbital rule for electron transport. Our results suggest that appropriate molecular design may indeed allow to construct highly sensitive molecular quantum rulers.
\end{abstract}


\maketitle 

\section{Introduction}\label{Intro}

Control of transport through molecular junctions by external parameters is crucial to improve our understanding of these nanosystems and to ultimately use them as functional units. The energy landscape of the transmission may be explored by electrostatic or electrochemical gating \cite{Koole:NanoLett2015,Gehring:NanoLett2017,Li:NatMater2019,Bai:NatureMater2019}, adjusting the alignment of molecular orbitals with respect to the Fermi energy. Other external control parameters may be of mechanical nature. The electrode displacement may change the molecular extent \cite{Li:NatComm2021}, leading to displacement of decks in $\pi$-stacks\cite{Wu:NatNano2008, Frisenda2016,Solomon:JChemPhysB2010,Solomon:FaradyDiscuss2014,nozakiMolecularOrbitalRule2017,stefaniLargeConductanceVariations2018,reznikovaSubstitutionPatternControlled2021,schosserMechanicalConductanceTunability2022,hsuMechanicalCompressionCofacial2022,vanderPoel:NatComm2024}, unfolding of a helicene spiral\cite{Guo:SciRep2015,Vacek:Nanoscale2015,deAra:JCPL2024}, or variations in the orientation of rings in ferrocene \cite{camarasa-gomezMechanicallyTunableQuantum2020}.

If the landscape of transmission versus energy and an in principle arbitrary control parameter is known, measurement of the electrical conductance informs conversely about the state of a molecule inside the junction. A molecule may then serve as a quantum sensor, detecting intrinsic or environmental interactions \cite{vanderPoel:NatComm2024}. 
For example, destructive quantum interferences can be tuned mechanically through the full gap between the highest occupied molecular orbital (HOMO) and the lowest unoccupied molecular orbital (LUMO) for $\pi$-stacked molecules\cite{nozakiMolecularOrbitalRule2017,stefaniLargeConductanceVariations2018,reznikovaSubstitutionPatternControlled2021,schosserMechanicalConductanceTunability2022,vanderPoel:NatComm2024}. By measuring the conductance upon changes in electrode displacement and counting destructive quantum interference dips, distances can in principle be determined in a similar way as counting constructive and destructive interferences upon changes in the length of a Fabry-Pérot interferometer in optics. If it were possible to generate grids of destructive quantum interferences in the HOMO-LUMO gap of a molecular junction, a molecular quantum ruler could be built to measure any kind of variation in the external control parameter electrically.

For such sensors to be highly sensitive, large changes in conductance are needed as a function of the external control. It has been shown that destructive quantum interference can cause changes of the conductance by orders of magnitude even at room temperature in specifically designed junctions that use mechanoelectrically sensitive molecules\cite{stefaniLargeConductanceVariations2018,reznikovaSubstitutionPatternControlled2021,schosserMechanicalConductanceTunability2022,vanderPoel:NatComm2024}. The conductance minima, caused by destructive interference, can then serve as reference points to quantify fluctuations\cite{vanderPoel:NatComm2024}.

It is therefore important to understand, which destructive quantum interference structures can be expected in molecular junctions. We consider specifically a four-level model and classify analytically and numerically the kind of interferences that occur. Multiple levels can lead to quite complex destructive interference structures inside the HOMO-LUMO gap, suggesting the possibility to realize molecular quantum rulers. In the following discussion we have mechanoelectrically sensitive molecular junctions in mind. The external control can thus be imagined to be the separation between two macroscopically large electrodes. But the external control parameter might be of any other kind such as a dihedral angle or a static magnetic field. 

The manuscript is structured as follows. We introduce in Sec.~\ref{sec:TransportDescription} the scattering formalism for the elastic coherent transport through molecular junctions. Subsequently, we present the well-established orbital rule for electron transport through molecules in Sec.~\ref{sec:OrbitalRule-ETransport}. We discuss in Sec.~\ref{sec:4LM}, how to reduce the description of transport to just four electronic levels. This four-level model is then studied in two versions, a simplified binary one in Sec.~\ref{sec:Binary-4LM} that can be treated analytically and a nonbinary one in Sec.~\ref{sec:Nonbinary-4LM} that has so many degrees of freedom that we restrict ourselves to special cases. We highlight crucial differences between the models, before concluding in Sec.~\ref{sec:Conclusions}. In the appendix, we explain further technical aspects such as the connection between the electronic transmission and the propagator, which we use as important concept to understand the interference of electron waves.

\section{Transport description}\label{sec:TransportDescription}

\begin{figure}
    \centering
    \includegraphics[width=0.8\linewidth]{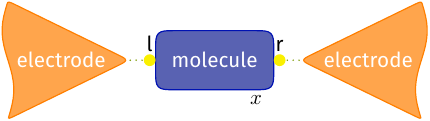}
    \caption{Sketch of a molecular junction consisting of a molecule attached to two electrodes. The energy levels and associated states of the molecule depend on a control parameter $x$. The yellow dots labeled `l' and `r' refer to atomic orbitals on the anchors of the molecule, e.g., sulfur atoms forming a bond to gold electrodes.}
    \label{fig:junction}
\end{figure}

We consider the linear conductance of a molecular junction, where a molecule is contacted by two metallic electrodes, as sketched in Fig.~\ref{fig:junction}. The size of the molecule and its environmental coupling should be such that the electronic transport through the molecule is coherent. According to Landauer-Büttiker scattering theory\cite{cuevasMolecularElectronics2017}, the conductance
\begin{equation}
    G(x) = \frac{2e^2}{h}\tau(E_\text{F},x)
\end{equation}
is then given by the  transmission $\tau$ at the Fermi energy $E_\text{F}$ up to the quantum of conductance $2e^2/h$. In the expression, we have additionally indicated the control parameter $x$, which modifies the transmission and consequently the conductance.

Contacts to left and right electrodes are assumed to be made by single atomic orbitals `l' and `r' on left and right anchoring atoms of the molecule, respectively; see Fig.~\ref{fig:junction}. Additionally, transport shall be off-resonant such that the transmission is typically much less than 1. Within the wide-band approximation for the electrodes, the transmission of electrons through the junction is then given by 
\begin{equation}
    \label{eq:transmission}
    \tau(E,x) \propto \vert G_{\text{lr}}^{\text{r}}(E,x)\vert^2,
\end{equation}
where 
\begin{equation}
    \label{eq:G0r-spectral}
    G_{\text{lr}}^{\text{r}}(E,x) = \sum_n\frac{c_{\text{l},n}(x)c_{\text{r},n}^*(x)}{E+i\eta-\epsilon_n(x)}=\sum_n\frac{\rho_n(x)}{E+i\eta-\epsilon_n(x)}
\end{equation}
is the spectral representation of the zeroth-order retarded Green's function. The sum runs over all molecular orbitals $n$ with energy $\epsilon_n$, which might be considered to be renormalized by the presence of the electrodes. The expansion coefficients of the $n$-th orbital at the anchoring molecular sites `l' and `r' are denoted by $c_{\text{l},n}$ and $c_{\text{r},n}$, respectively, and the star indicates the complex conjugate. The product of the orbital expansion coefficients is abbreviated as the residue $\rho_{n}=c_{\text{l},n}c^*_{\text{r},n}$. The imaginary part $\eta$ can be viewed to arise from the finite lifetime acquired by the molecule-electrode coupling. For numerical purposes, we choose $\eta=10^{-4}\Delta$ to avoid divergences of the Green's function $G_{\text{lr}}^{\text{r}}$, with $\Delta$ being basically the size of the HOMO-LUMO gap. For its precise definition within the four-level model, see Eqs.~\eqref{eq:epsH_x} and \eqref{eq:epsL_x}. Within the analytical treatment, $\eta$ will be set to zero, and to indicate this, we will then drop the superscript `r' and denote the propagator as $G_{\text{lr}}$. While a finite value of $\eta$ will lead to a broadening, the features in the transmission will be robust as long as $\eta \ll \Delta$.
As far as the real part of the self-energies is concerned, we assume that the energies $\epsilon_n$ refer to the dressed levels of a molecule attached to two leads. We refer the reader to Sec.~\ref{sec:TransmissionPropagator-connection-appendix} for further explanations of approximations involved.

In the remainder of this paper, we will not be interested in absolute values of the conductance and will therefore omit proportionality constants. Our main interest will thus be in the squared propagator, i.e.\ the right-hand-side of Eq.~\eqref{eq:transmission}, together with the spectral decomposition, Eq.~\eqref{eq:G0r-spectral}. In particular, our focus will be on zeros of the transmission due to the interference of contributions arising from different molecular orbitals.

\section{Orbital rule for electron transport through molecules}\label{sec:OrbitalRule-ETransport}

\begin{figure}
    \begin{center}
        \includegraphics[width=0.8\columnwidth]{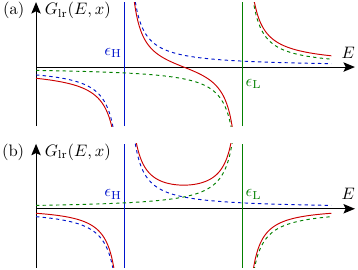}
    \end{center}
    \caption{Illustration of the orbital rule for electron transport through molecules with regard to
    the interference of contributions of the frontier orbitals HOMO and LUMO. The blue and green dashed curves represent individual Green's functions for the HOMO and the LUMO level, respectively, while the red solid curve represents their sum. In panel (a) the signs of the residues $\rho_{\text{H}}$ and $\rho_{\text{L}}$ are the same, while they differ in panel (b). 
    }
    \label{fig:yoshizawa_rule}
\end{figure}

An orbital rule for the electron transport through junctions made from aromatic molecules has been derived by Yoshizawa et al.~\cite{Yoshizawa:JACS2008,yoshizawaOrbitalRuleElectron2012} by taking into account the two frontier orbitals HOMO and LUMO, abbreviated in the following by H and L, respectively. Two aspects determine the conductance of the molecular junction. The contribution of each orbital depends on the magnitude of the coefficients $c_{\text{l},n}$ and $c_{\text{r},n}$ at the terminal sites of the molecule, to which the metallic electrodes are coupled, see Fig.~\ref{fig:junction}. In this respect, the symmetry properties of the molecular orbital wavefunctions play a crucial role. In addition, interference effects between the different molecular orbitals are relevant. Taking the expansion coefficients to be real, the relative sign of the residues $\rho_{n}=c_{\text{l},n}c_{\text{r},n}$ for different orbitals $n$ is decisive. This is particularly true for the suppression of the transmission at positions between the energies of the two frontier orbitals. A destructive interference of the contributions of the two frontier orbitals can occur if the signs of $\rho_\text{H}$ and $\rho_\text{L}$ agree, as illustrated in Fig.~\ref{fig:yoshizawa_rule}a. In contrast, for the case of differing signs depicted in Fig.~\ref{fig:yoshizawa_rule}b, the conductance cannot vanish in the gap between HOMO and LUMO. 

\section{Reduction to the four-level model}\label{sec:4LM}
In the following, we will focus on the case of four orbitals in the sum of Eq.~\eqref{eq:G0r-spectral} and explore the emerging rich structure of zeros of the transmission inside the HOMO-LUMO gap. The choice of four levels is motivated by the architecture of mechanoelectrically sensitive molecules.\cite{nozakiMolecularOrbitalRule2017,stefaniLargeConductanceVariations2018,reznikovaSubstitutionPatternControlled2021,schosserMechanicalConductanceTunability2022,hsuMechanicalCompressionCofacial2022,vanderPoel:NatComm2024} They typically consist of two identical conjugated molecular decks that can be mechanically shifted against each other. The weak tunnel-splitting of the HOMO and LUMO of each deck, arising from the coupling of $\pi$ electrons, leads to four relevant molecular states. Their energetic order can be varied, since the sign of the tunnel-splitting of HOMOs and LUMOs depends on the displacement of the molecular decks.\cite{nozakiMolecularOrbitalRule2017,stefaniLargeConductanceVariations2018} Further energy levels can be neglected, provided that the applied voltage and the temperature are sufficiently small. Below, we will collectively refer to HOMO-1 and HOMO as HOMOs and correspondingly for LUMOs. We will also assume that the residues $\rho_n$ of all levels are nonzero, as otherwise their number would effectively be less than four.

In the following, we will study the transmission as a function of energy $E$ and a control parameter $x$. While the Fermi energy can be varied by applying a gate voltage, $x$ can be related, e.g., to a mechanical force applied to the molecule. Alternatively, one can imagine applying a torsion to the molecule or modifying some other relevant degree of freedom, which influences the energy levels and states involved. The dependence of the transmission on $E$ and $x$ can thus, in principle, be mapped experimentally.

When a molecule is manipulated, both the energies of the molecular orbitals as well as the orbital wavefunctions themselves will change. Variations of the transmission in Eq.~\eqref{eq:transmission} as a function of the control parameter $x$ may thus result from modifications of the $\epsilon_n$ as well as the residues $\rho_n$, as indicated in Eq.~\eqref{eq:G0r-spectral}. We will from now on restrict the sum over $n$ in Eq.~\eqref{eq:G0r-spectral} to the two highest occupied levels, denoted $\text{H},+$ and $\text{H},-$, and the two lowest unoccupied levels, denoted $\text{L},+$ and $\text{L},-$. For numerical purposes, we will vary the energy levels as a function of the control parameter $x$ according to
\begin{equation}
    \epsilon_{\text{H},\pm}(x) = \Delta\left[-\frac{1}{2} \pm a_\text{H}(x)\right]
    \label{eq:epsH_x}
\end{equation}
for the two highest occupied orbitals and
\begin{equation}
    \epsilon_{\text{L},\pm}(x) = \Delta\left[\frac{1}{2} \pm a_\text{L}(x)\right]
    \label{eq:epsL_x}
\end{equation}
for the two lowest unoccupied orbitals. We choose $a_\text{L}+a_\text{H} < 1$ to ensure that the HOMO-LUMO gap $\Delta(1-a_\text{H}-a_\text{L})$ is always positive. The zero of energy always lies in the middle of the HOMO-LUMO gap. For convenience, we vary the energies periodically as a function of the control parameter as
\begin{equation}
    a_\text{H}(x)=\delta_\text{H}\cos(2\pi x)\label{eq:aH_x_periodic}
\end{equation}
and 
\begin{equation}
    a_\text{L}(x)=\delta_\text{L}\cos(2\pi\alpha x)\label{eq:aL_x_periodic}
\end{equation}
with the dimensionless coefficients $\delta_\text{H}, \delta_\text{L}$ and $\alpha$. Due to the lack of a detailed molecular model, we will in the following only shortly numerically discuss the effect of $x$-dependent variations of $\rho_n(x)$ on the transmission in Sec.~\ref{sec:Nonbinary-4LM}. In the largest part of the following discussion, the residues will be assumed to be independent of $x$.

The dependencies \eqref{eq:aH_x_periodic} and \eqref{eq:aL_x_periodic} are not necessarily meant to model a realistic dependence of the molecular energy levels on the control parameter, even though an oscillating behavior can be observed in $\pi$-stacks.\cite{nozakiMolecularOrbitalRule2017,stefaniLargeConductanceVariations2018,Schosser:PhD2022} They rather allow us to study the appearance of destructive quantum interference for different sequences of the signs of the residues $\rho_n$ and as a function of the varying energy differences between the two HOMO levels and the two LUMO levels, respectively.

To investigate destructive quantum interferences analytically, it is convenient to set $\eta=0$ in the Green's function \eqref{eq:G0r-spectral}. Then, instead of searching for transmission valleys, it is sufficient to determine zeros of $G_\text{lr}$. In realistic situations, the imaginary part of $\eta$ will not vanish, leading to a broadening of transmission valleys and a somewhat incomplete destructive interference. Bringing $G_\text{lr}$ to the common denominator, the search for zeros of $\tau$ reduces to determining the zeros of the resulting numerator
\begin{equation}
    f(E,x)=\sum_n \rho_n(x) \prod_{i\neq n}\big(E-\epsilon_i(x)\big)\,.\label{eq:f_numeratorG}
\end{equation}
Since for four levels $n$ the product runs over three factors, the numerator $f(E,x)$ will typically be a third-order polynomial in $E$. However, if the sum of the residues vanishes, the numerator will simplify to a quadratic polynomial.

\section{Binary four-level model}\label{sec:Binary-4LM}

\begin{figure*}[t!]
    \begin{center}
        \includegraphics[width=\linewidth]{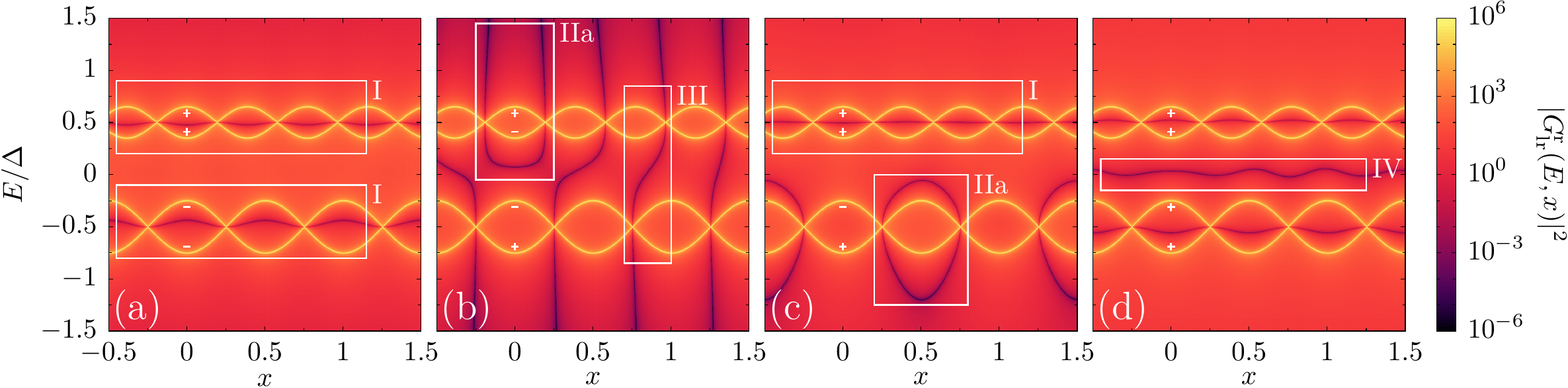}
    \end{center}
    \caption{The square modulus of the propagator as a function of the energy $E$ and the control parameter $x$ is shown for the four basic sign combinations of the residues in a binary four-level model: (a) \texttt{--++}, (b) \texttt{+--+}, (c) \texttt{+-++} and (d) \texttt{++++}. The signs are listed in energetically ascending order and are indicated at $x=0$ in each panel. The different structure types of the transmission valleys are marked by numbered white boxes. The parameters appearing in the level energies as a function of the control parameter $x$, see Eqs.~\eqref{eq:epsH_x} to \eqref{eq:aL_x_periodic}, are chosen as $\delta_\text{H}=0.25$, $\delta_\text{L}=0.15$ and $\alpha=1.3$.}
    \label{fig:4level-DQI-types}
\end{figure*}

In this section, we will analytically study destructive quantum interference in a simplified four-level model as introduced in Ref.~\onlinecite{yoshizawaOrbitalRuleElectron2012}, where all residues $\rho_n$ have the same absolute value. Similar models were used by other authors.\cite{liOrbitalViewsElectronTransport2012, nozakiMolecularOrbitalRule2017, stefaniLargeConductanceVariations2018, schosserMechanicalConductanceTunability2022} Without loss of generality, we can set each residue $\rho_n$ to one of the two values $1$ and $-1$, as done in Ref.~\onlinecite{stefaniLargeConductanceVariations2018}, and therefore refer to the model as binary four-level model.

Even though the binary four-level model in principle allows for sixteen different combinations of signs, it is possible to restrict our considerations to four cases. Clearly, multiplying all residues by $-1$ will not change the structure of the transmission valleys. Furthermore, exchanging $E$ by $-E$ will not lead to new structures. Finally, we allow $a_\text{H}$ and $a_\text{L}$ to have either sign. For our numerical results, these sign changes are ensured by the dependencies \eqref{eq:epsH_x} to \eqref{eq:aL_x_periodic} as a function of the control parameter $x$. As a consequence, it is sufficient to consider the four sign combinations \texttt{--++}, \texttt{+--+}, \texttt{+-++} and \texttt{++++}, which we assign to the residues of the levels for ascending energy at $x=0$. According to the remark at the end of the previous section, the first two cases will lead to a quadratic polynomial $f(E,x)$ in Eq.~\eqref{eq:f_numeratorG} as a function of $E$, while the last two cases result in a cubic polynomial.

We start our analytical investigation with the simplest case, where the signs of the residues for the levels with increasing energy are given by \texttt{--++}. From Fig.~\ref{fig:yoshizawa_rule}a we know that between two levels, for which the sign of the residues agree, there necessarily exists an odd number of zeros. As there are two such pairs, namely the two LUMOs as well as the two HOMOs, there will be a destructive quantum interference minimum between each of the two pairs. Since $f(E,x)$ is given by a quadratic polynomial, we have already identified all destructive quantum interferences. This situation is visualized in Fig.~\ref{fig:4level-DQI-types}a. Degeneracy points of the energies $\epsilon_{\text{H},\pm}$ and $\epsilon_{\text{L},\pm}$ are thus connected horizontally by a transmission valley, and we refer to such a horizontal valley either between the two HOMOs or the two LUMOs as type~I.

The second sign combination, leading to a quadratic polynomial for $f(E,x)$, is \texttt{+--+}. For the choice of energies \eqref{eq:epsH_x} and \eqref{eq:epsL_x} as a function of the control parameter $x$, we find regions, where the signs of the residues for the HOMO and LUMO agree. This is the case for example around $x=0$. On the other hand, the LUMOs go through a degeneracy point at $x=1/4\alpha\approx0.19$ in Fig.~\ref{fig:4level-DQI-types}b, and the sequence of the signs of the residues is subsequently inverted. As long as the HOMOs in Fig.~\ref{fig:4level-DQI-types}b do not go through a degeneracy, we no longer find two neighboring levels with equal signs of the residues.

In order to determine whether any zeros of $f(E,x)$ exist in the different regions of the control parameter, we take a closer look at the corresponding quadratic polynomial
\begin{equation}
    E^2+E\frac{a_\text{L}(x)+a_\text{H}(x)}{a_\text{L}(x)-a_\text{H}(x)}+\frac{1}{4}-a_\text{H}(x)a_\text{L}(x) \overset{!}{=}0\,.
    \label{eq:quadratic_condition}
\end{equation}
The signs of the residues in the HOMO pair or the LUMO pair can be interchanged by reversing the sign of $a_\text{H}$ or $a_\text{L}$, respectively, see Eqs.~\eqref{eq:aH_x_periodic} and \eqref{eq:aL_x_periodic}. It is straightforward to determine the corresponding discriminant
\begin{equation}
    D(x) = 4a_\text{H}(x)a_\text{L}(x)\left(1+\frac{1}{(a_\text{H}(x)-a_\text{L}(x))^2}\right)\,,
\end{equation}
which is positive provided $a_\text{H}$ and $a_\text{L}$ have the same sign, i.e.\ if the signs of the residues are either \texttt{+--+} or \texttt{-++-}. On the other hand, for the combinations of signs \texttt{-+-+} or \texttt{+-+-}, the discriminant is negative and no zeros exist for real energies $E$. In Fig.~\ref{fig:4level-DQI-types}b, we thus find regions of $x$, where two zeros exist as a function of energy or where no zeros exist at all. Note that the second zero may not be visible in Fig.~\ref{fig:4level-DQI-types}b, because it lies outside the range of energies shown.

\begin{figure}[t!]
    \begin{center}
        \includegraphics[width=0.8\linewidth]{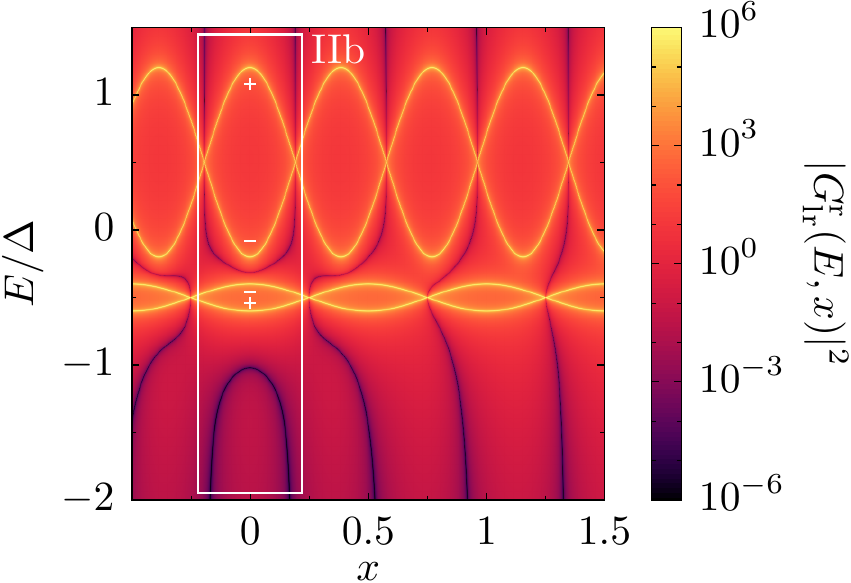}
    \end{center}
    \caption{The square modulus of the propagator as a function of the energy $E$ and the control parameter $x$ is shown for the sign combination \texttt{+--+} of the residues in the binary four-level model with $\delta_\text{H}=0.1$, $\delta_\text{L}=0.7$ and $\alpha=1.3$. The two parts of the transmission valley structure, referred to as type IIb, are connected at infinite energy.}
    \label{fig:u_valley}
\end{figure}

In order to understand the existence of zeros above the highest level and below the lowest level, it is sufficient to consider the asymptotic behavior of $G_\text{lr}$. For the signs \texttt{+--+} of the residues, one finds
\begin{equation}
  G_\text{lr}(E,x) \sim \frac{2(a_\text{L}(x)-a_\text{H}(x))}{E^2}\qquad\text{for $\vert E\vert\to\infty$}\,.
  \label{eq:asymptotics}
\end{equation}
For the case $a_\text{H} > a_\text{L}$, see Fig.~\ref{fig:4level-DQI-types}b,
$G_\text{lr}$ thus approaches zero from below for large energies. On the other hand, for energies slightly exceeding the energy of the uppermost level, $G_\text{lr}$ is positive. Therefore, a zero has to exist somewhere above the uppermost level. For the parameters chosen in Fig.~\ref{fig:4level-DQI-types}b, the two branches of the transmission valley in the vicinity of $x=0$ join at an energy $E\approx 3.9\Delta$ outside of the energy range displayed here. We designate such a loop structure of the transmission valley as type~IIa.

If, on the other hand, $a_\text{H} < a_\text{L}$, the asymptotic behavior \eqref{eq:asymptotics} changes sign and, following the same line of reasoning, a zero occurs at an energy below the lowest level. For $a_\text{L}\approx a_\text{H}$, one solution of Eq.~\eqref{eq:quadratic_condition} is found at $E=(a_\text{H}+a_\text{L})/(a_\text{H}-a_\text{L})$. If $a_\text{L}$ changes from values smaller than $a_\text{H}$ to values larger than $a_\text{H}$, the zero moves to infinite energy and reappears at negative infinite energy. This behavior is illustrated in Fig.~\ref{fig:u_valley}, and the transmission valley is classified as type~IIb. Of course, this behavior is specific for the four-level model and will be perturbed by the presence of other energy levels. 

As a result of this discussion, we see that the destructive quantum interferences for signs of the residues of the type \texttt{+--+} are never found between the pair of HOMOs or the pair of LUMOs. Degeneracies are connected by transmission valleys running outside the HOMO or LUMO pairs. If two degeneracies either in the pair of LUMOs or in the pair of HOMOs are connected, we obtain a structure of type~II which we already discussed. In addition, it is possible that a transmission valley connects a degeneracy in the HOMO pair with a degeneracy in the LUMO pair, thus crossing the entire HOMO-LUMO gap. Such structures are also shown in Fig.~\ref{fig:4level-DQI-types}b and designated as type~III. The upper and lower ends join at infinite energy if we identify positive and negative infinity.

\begin{figure}[t!]
    \centering
    \includegraphics[width=0.6\linewidth]{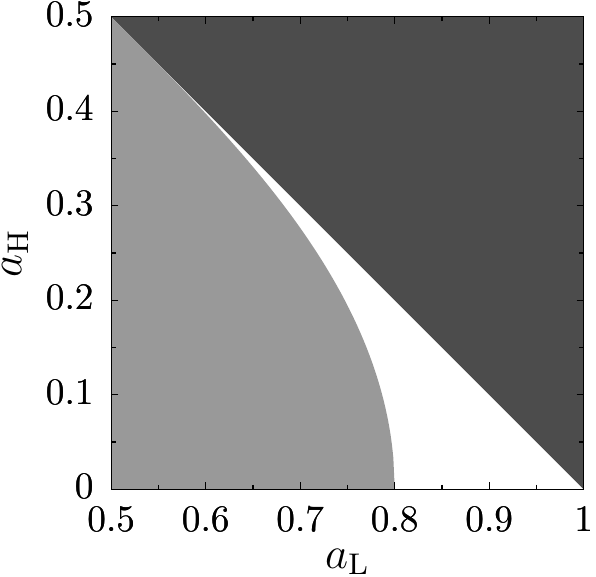}
    \caption{In the white region, a transmission valley of type~V occurs. The region excluded by the first and second inequality in Eq.~\eqref{eq:condition_v} is marked by the light grey and dark grey areas, respectively.}
    \label{fig:type_v_condition}
\end{figure}

The remaining two cases \texttt{+-++} and \texttt{++++}, shown in Fig.~\ref{fig:4level-DQI-types}c and Fig.~\ref{fig:4level-DQI-types}d, respectively, lead to cubic polynomials for $f(E,x)$ as a function of $E$. For the case \texttt{++++}, depicted in Fig.~\ref{fig:4level-DQI-types}d, all residues have the same sign. Therefore, three transmission valleys occur, one valley between each successive pair of levels. We thus have two transmission valleys of type~I. In addition, a horizontal transmission valley appears in the HOMO-LUMO gap, which we refer to as type~IV.

The other case \texttt{+-++} requires a more detailed analysis. Because the signs of the residues of the two LUMOs agree, we find a transmission valley of type~I between the two LUMOs. 

In regions of $x$, where the signs of the residues agree for the HOMO and LUMO, e.g.\ around $x=0.5$ in Fig.~\ref{fig:4level-DQI-types}c, a destructive quantum interference appears inside the HOMO-LUMO gap. Then, for a similar reason as for the sign combination \texttt{+--+}, a third transmission valley occurs simultaneously below the four levels. In this situation with the sign order of residues \texttt{-+++}, the lowest level with a negative sign of the residue makes a large positive contribution to $G_\text{lr}$ for energies below the lowest level. On the other hand, asymptotically for $E\to-\infty$, $G_\text{lr} \sim 2/E$, thus approaching zero from below. As a consequence, a transmission valley occurs below the lowest level. The resulting loop structure of the transmission valley is of type~IIa. Originally, we had introduced type~II structures as solutions of the quadratic equation \eqref{eq:quadratic_condition}. However, dividing out a linear factor resulting from the zero of type~III, we obtain a quadratic equation for which our previous line of reasoning applies.

In regions, where the HOMO and LUMO have residues of different sign such as around $x=0$ in Fig.~\ref{fig:4level-DQI-types}c, there will be no transmission valley in the HOMO-LUMO gap, because the contribution of the LUMOs in the gap is negative and the same holds true for the sum of the contributions of the HOMOs. Destructive quantum interference below the lowest level is also excluded. However, one can imagine that there occur two transmission valleys between the two HOMOs. This is not the case for the parameters used in Fig.~\ref{fig:4level-DQI-types}c. However, reinterpreting $\epsilon_\text{H}$ and $\epsilon_\text{L}$ in Fig.~\ref{fig:yoshizawa_rule}b as the energies of two HOMOs, the negative contribution of the LUMOs could pull the red curve below zero, resulting in two zeros.

\begin{figure}[!t]
    \centering
    \includegraphics[width=0.8\linewidth]{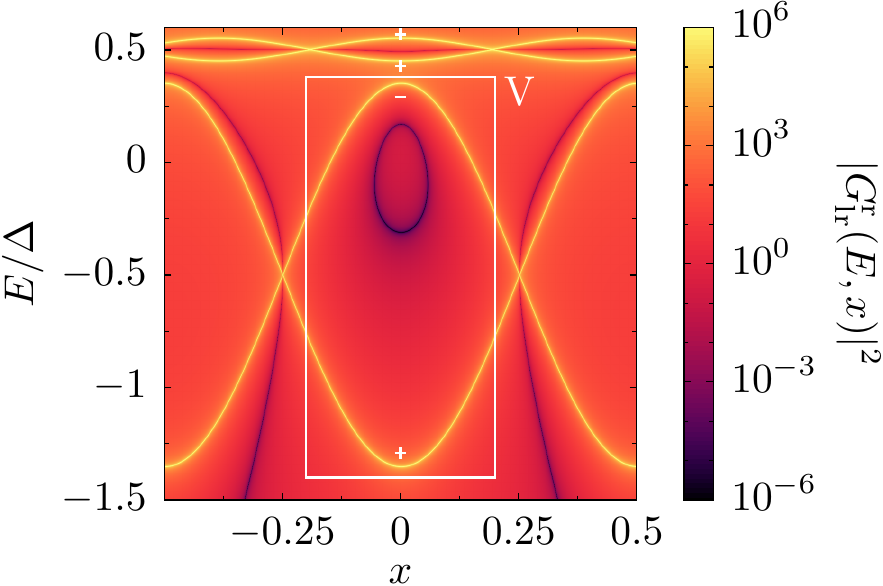}
    \caption{A transmission valley of type~V is shown for the sign combination \texttt{+-++}, using the parameters $\delta_\text{H}=0.05$, $\delta_\text{L}=0.85$ and $\alpha=1.3$.}
    \label{fig:innerloop}
\end{figure}

In order to explore this scenario, we apply a suitable shift of the energy to bring the cubic polynomial for $f(E,x)$ into its reduced form
\begin{equation}
    E^3+p(x)E+q(x) = 0
\end{equation}
with
\begin{equation}
    p(x) = -\frac{(2a_\text{L}(x)-1)^3}{3}
    \label{eq:cubic_p}
\end{equation}
and
\begin{multline}
    q(x) = -\frac{11}{27}\left(a_\text{L}(x)-\frac{1}{2}\right)^3+\frac{1}{2}\left(a_\text{L}(x)-\frac{1}{2}\right)^2\\
        +\frac{1}{4}(a_\text{L}(x)-1)+a_\text{L}(x)a_\text{H}^2(x)\,.
    \label{eq:cubic_q}
\end{multline}
Three real zeros of $f(E,x)$ and thus two transmission valleys between the pair of HOMOs occur, if the discriminant $D=(p/3)^3+(q/2)^2$ of the cubic equation is negative. Inserting expressions \eqref{eq:cubic_p} and \eqref{eq:cubic_q} into $D$ and solving for $a_\text{H}$, one finds that a total of three transmission valleys can occur for the sign combination \texttt{+-++}, if the condition
\begin{multline}
    (a_\text{L}(x)+1)\sqrt{\frac{4-5a_\text{L}(x)}{27a_\text{L}(x)}} < \vert a_\text{H}(x)\vert < 1-a_\text{L}(x)\\
    \quad\text{for $a_\text{L}(x) > \frac{1}{2}$}
    \label{eq:condition_v}
\end{multline}
is fulfilled. The constraint imposed on $a_\text{H}$ by the left inequality only applies for $a_\text{L} < 4/5$. The right inequality not only results from the requirement $D < 0$ but also ensures that there exists a nonvanishing HOMO-LUMO gap. In Fig.~\ref{fig:type_v_condition}, the parameters, for which the inequalities in Eq.~\eqref{eq:condition_v} are fulfilled, are indicated by the white area. The areas shown in light grey and dark grey are excluded by the first and second inequality in Eq.~\eqref{eq:condition_v}, respectively. As an example, Fig.~\ref{fig:innerloop} shows the appearance of a transmission valley of type~V, forming an inner loop between the two HOMOs.

For the binary four-level model we find that the occurrence of transmission valleys follows the orbital rule for electron transport through molecules\cite{Yoshizawa:JACS2008,yoshizawaOrbitalRuleElectron2012}: Inside the HOMO-LUMO gap, either no or one destructive quantum interference emerges. Destructive quantum interference occurs for energies between two levels, for which the signs of the residues agree. However, it turns out that under certain circumstances, a valley of type~V can occur between the two HOMOs or between the two LUMOs, even though the signs of the respective residues differ.

\section{Nonbinary four-level model}\label{sec:Nonbinary-4LM}

\begin{figure}
    \centering
    \includegraphics[width=0.8\linewidth]{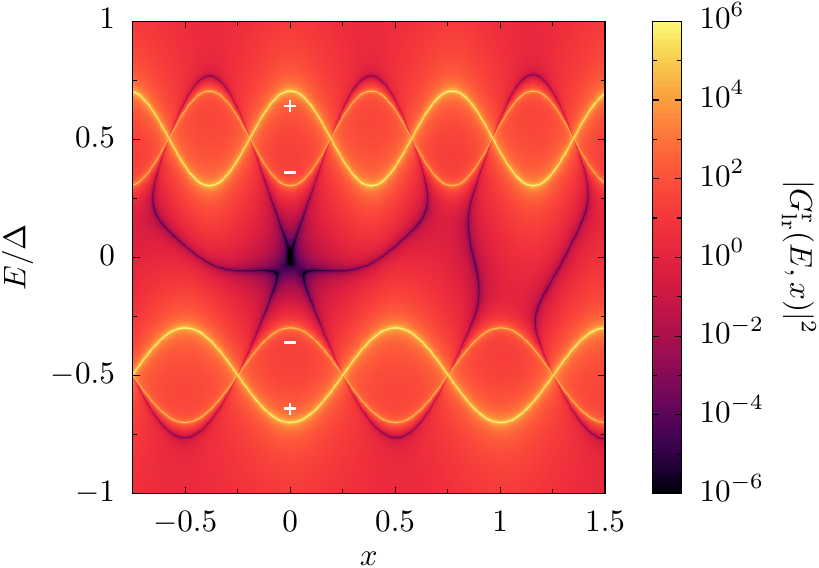}
    \caption{The square modulus of the propagator in the nonbinary four-level model as a function of the energy $E$  shows up to three destructive quantum interferences in the HOMO-LUMO gap at a given control parameter $x$. The energy levels are given by the relations \eqref{eq:epsH_x} to \eqref{eq:aL_x_periodic} with $\delta_\text{H}=\delta_\text{L}=\delta=0.2$ and $\alpha=1.3$. At $x=0$, the residues were chosen as $\rho_{\text{H},-}=\rho_{\text{L},+}=1$ and $\rho_{\text{H},+}=\rho_{\text{L},-}=-\rho_0$.}
    \label{fig:loop}
\end{figure}

While the binary four-level model, treated in the previous section, lends itself for an analytical treatment, the choice of residues $\pm1$ is very special. But then, the full four-level model features a fairly large parameter space, which complicates an analysis. Therefore, we focus on the destructive quantum interference characteristics in the HOMO-LUMO gap. As we have seen, the number of destructive quantum interferences in the gap is limited to one in the case of the binary four-level model. Figure~\ref{fig:loop} shows that complex transmission valley structures can appear in the HOMO-LUMO gap in the nonbinary four-level model. In particular, for four levels, it is possible that three destructive quantum interferences exist inside the gap at a given $x$.

\begin{figure}
    \begin{center}
        \includegraphics[width=0.8\columnwidth]{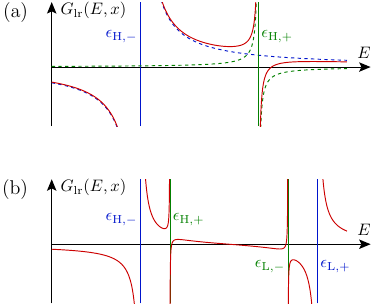}
    \end{center}
    \caption{(a) Sketch of the propagator for two HOMOs with residues of different sign and different absolute value. A zero in the propagator for energies above the higher level can occur, if the magnitude of the residue of this level is the smaller one. (b) With a mirrored scenario for the two LUMOs, three zeros are found in the HOMO-LUMO gap.}
    \label{fig:additional_zero}
\end{figure}

In order to avoid the complexity of the full four-level model, we consider some simplifications. As done in Fig.~\ref{fig:loop}, we set the residues of the HOMO and LUMO at $x=0$ to $-\rho_0$, while the residues of the other orbitals are set to one. As the signs of the residues of HOMO and LUMO agree in the region around $x=0$, we expect either one or three transmission valleys in the HOMO-LUMO gap. The existence of more than one destructive quantum interference can be explained by an argument, which resembles the reasoning for a transmission valley above or below the four levels, given in the previous section. The scenario describes for example the situation, realized in Fig.~\ref{fig:loop} for $-1/4\alpha<x<1/4\alpha$. As depicted in Fig.~\ref{fig:additional_zero}a, the residues of the two HOMO levels have different signs, but the absolute value of the residue of the upper level $\text{H},+$ is smaller. Close to the upper level, the negative contribution of this level dominates, while the lower HOMO level $\text{H},-$ with its larger positive residue is decisive at larger energies. Therefore, a zero in the propagator occurs. If the mirror image of this scenario holds for the two LUMOs, one finds three zeros, as shown in Fig.~\ref{fig:additional_zero}b. 

Let us now try to obtain a better analytical understanding of the situation, shown in Fig.~\ref{fig:loop}. We therefore study the case, where the residues of the HOMO and LUMO are given by $-\rho < 0$, while the outer two orbitals have residues equal to one. Furthermore, we assume that the level pairs are equally spaced at energies $\epsilon_{\text{H},\pm}=-\Delta(\frac{1}{2}\pm a)$ and $\epsilon_{\text{L},\pm}=\Delta(\frac{1}{2}\pm a)$. The zeros of the propagator are then given by solutions of the cubic equation
\begin{equation}
    E\left[E^2(1-\rho)+\rho\left(\frac{1}{2}+a\right)^2-\left(a-\frac{1}{2}\right)^2\right] = 0\,.
\end{equation}
For $\rho_0=(2a-1)^2/(2a+1)^2$, the three zeros will degenerate to yield a single zero, as can be seen in Fig.~\ref{fig:loop} at $x=0$. For $\rho < \rho_0$, the degeneracy is lifted and a total of three zeros emerges. Note that the deviation of $\rho$ from $\rho_0$ in Fig.~\ref{fig:loop} is achieved by the different dependence of $a_\text{H}$ and $a_\text{L}$ on $x$. For non-vanishing values of $\rho$ close to zero, the two additional zeros move towards the energies of the HOMO and LUMO, see also Fig.~\ref{fig:additional_zero}b. 

\begin{figure}[t!]
    \centering
    \includegraphics[width=0.8\linewidth]{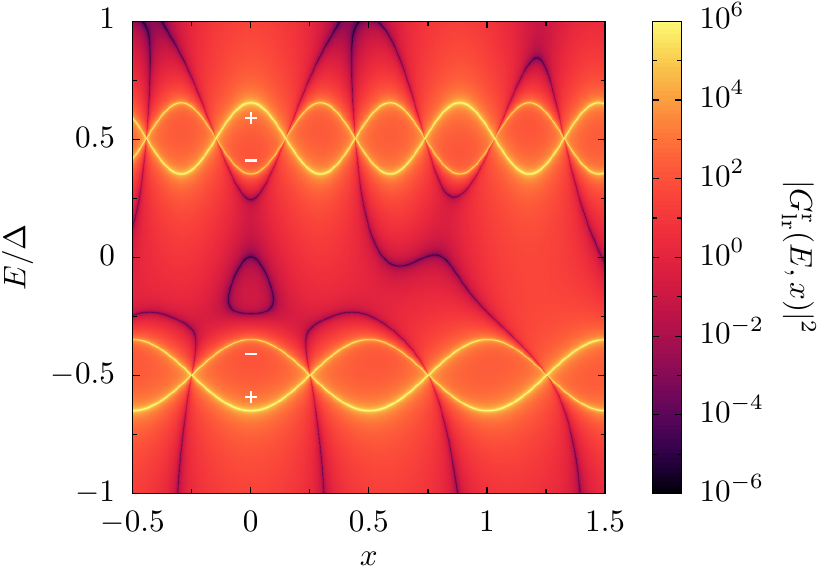}
    \caption{Square modulus of the propagator for a four-level model with parameters $\delta_\text{H}=\delta_\text{L}=0.15$ and $\alpha=1.7$. The residues depend on the control parameter as $\rho_{\text{H},\pm}(x)=0.7\cos(\pi\beta_\text{H} x)\mp1$ and $\rho_{\text{L},\pm}(x)=0.7\cos(\pi\beta_\text{L} x)\pm1$ with $\beta_\text{H}=1.7$ and $\beta_\text{L}=1$.}
    \label{fig:example_advanced_model}
\end{figure}

Inspecting Fig.~\ref{fig:loop} closely, we notice another interesting feature of the nonbinary four-level model. Zeros also exist in intervals of the control parameter, where the signs of the HOMO and the LUMO differ. Such regions are for example $0.57\lesssim x \lesssim 0.65$ and $1.16 \lesssim x < 1.25$. Although we found a possibility to induce two zeros between the HOMO-1 and HOMO in the binary model, see Fig.~\ref{fig:innerloop}, this time the zeros appear in the HOMO-LUMO gap. This behavior is in contrast to the orbital rule for electron transport through molecules valid for a two-level model, nor can it occur in the binary four-level model, as shown in the previous section. 

In contrast to the binary four-level model, where at most one destructive quantum interference can occur inside the HOMO-LUMO gap at a given $x$, the nonbinary four-level model in general allows for any number of transmission valleys in the HOMO-LUMO gap between zero and three, depending on the choice of the values of residues. While the parameter space is large and its full exploration is beyond the scope of this paper, we can at least say that for three destructive quantum interferences to arise in the HOMO-LUMO gap at a given $x$, the signs of the residues of HOMO and LUMO must be the same and the signs of the residues within the two HOMOs and within the two LUMOs must differ. Otherwise, transmission valleys of type~I would exist outside the HOMO-LUMO gap and the number of zeros inside the HOMO-LUMO gap would be at most two. 

So far, we have taken the residues to be independent of the control parameter $x$. In general, this will not be the case. Without an appropriate model for a concrete situation, we limit ourselves to present one example, which exhibits an interesting structure in the HOMO-LUMO gap. In Fig.~\ref{fig:example_advanced_model}, we have chosen residues to oscillate as a function of the control parameter. The choice of parameters is given in the figure caption. In this specific example, we find a transmission valley with a loop structure as a function of $E$ and $x$ near $x=0$. Furthermore, a transmission valley of type~III connects degeneracies in the HOMOs and the LUMOs. However, it exhibits a non-monotonic behavior as a function of $x$ and does not connect the two closest degeneracies.

\section{Conclusions}\label{sec:Conclusions}
We have presented a comprehensive analytical study of the structure of transmission valleys appearing in the binary four-level model as a function of energy $E$ and a control parameter $x$. In agreement with previous work, it is found that within the HOMO-LUMO gap at most one destructive quantum interference can exist at a fixed value of $x$. A necessary condition is that the signs of the residues of the two frontier levels agree.

Within a nonbinary four-level model, we have kept the residues of a HOMO- and a LUMO-related level equal, but different from the residues of the two other levels. Then, it was found that up to three destructive quantum interferences can exist inside the HOMO-LUMO gap at a given $x$, and a condition on the residues was given for this case. Finally, we have presented an example with residues depending on the control parameter, where even more interesting structures appear for the transmission in the HOMO-LUMO gap. 

The global structure of the transmission valleys discussed in this paper will of course be affected by additional levels not accounted for in the four-level model. If we generalize to $n$ molecular orbitals in the Green's function in Eq.~\eqref{eq:G0r-spectral}, we might obtain up to $n-1$ zeros inside the HOMO-LUMO gap. However, one would expect that energetically distant levels will require increasingly larger residues to affect the transmission in the HOMO-LUMO gap. Nevertheless, at least in principle, much more complicated multiple destructive quantum interferences as a function of energy $E$ can then arise at a fixed $x$ than what we have discussed here.

Complex destructive quantum interference structures could be used to build precise molecular quantum rulers to electrically measure both energetic and control parameter separations, e.g.\ changes in electrode displacement, using the molecule inside a molecular junction as a quantum sensor. In this way, much might be learned about fluctuations that single-molecule junctions are subject to. \cite{vanderPoel:NatComm2024}

It is presently not clear in all the cases, how a molecule would need to look like that shows the desired transmission properties. Destructive quantum interferences that cross the entire HOMO-LUMO gap (type III in Fig.~\ref{fig:4level-DQI-types}) have been reported for $\pi$-stacked molecules early on.\cite{nozakiMolecularOrbitalRule2017,stefaniLargeConductanceVariations2018} In simulations of anthracene double-deckers with various anchor group attachment points using extended Hückel parameters, we find between zero and two destructive quantum interferences inside the HOMO-LUMO gap for selected electrode separations, and some destructive interference valleys are loop-like (type II in Fig.~\ref{fig:4level-DQI-types}). The choice of appropriate intermolecular bridges\cite{schosserMechanicalConductanceTunability2022} or the attachment points of anchoring groups\cite{hsuMechanicalCompressionCofacial2022} may offer ways to steer the sliding motion of the $\pi$-electron systems of two molecular decks. An unfolding, as in helicene spirals, may additionally allow for changing orbital orientation upon stretching, offering possibilities to vary residue sizes. We hope that the predicted quantum interference types will be observed in the future. The finding of suitable molecular structures represents an exciting challenge for talented synthetic chemists and experimentalists.

\begin{acknowledgments}
We thank Abraham Nitzan for his many inspiring scientific contributions to molecular electronics and stimulating discussions on this topic. Abe, we wish you all the best to your birthday! Stay healthy and active in the years to come, and enjoy life scientifically and privately!

In addition, we acknowledge discussions on fascinating mechanoelectrically sensitive molecules in regular meetings with Herre van der Zant, Marcel Mayor, Nicolas Agra\"it and their groups. The present work was motivated by initial numerical studies of different forms of destructive quantum interference by Werner Schosser \cite{Schosser:PhD2022}. L.K.\ thanks Matthias Blaschke for the excellent supervision during his Bachelor thesis. L.K.\ and F.P.\ gratefully acknowledge funding by the Collaborative Research Center 1585, Project C02 of the German Research Foundation (grant number 492723217).
\end{acknowledgments}

\appendix

\section{Connection between transmission and zeroth order Green's function}\label{sec:TransmissionPropagator-connection-appendix}

In order to better understand the assumptions behind Eqs.~\eqref{eq:transmission} and \eqref{eq:G0r-spectral}, let us shortly sketch a derivation. The electronic transmission of a single-molecule junction can be expressed by the Landauer formula\cite{cuevasMolecularElectronics2017,paulyPhasecoherentElectronTransport2007} as 
\begin{equation}
    \tau(E,x) = \mathrm{Tr}\left[\mathbf{\Gamma}_\text{L}(E,x) \mathbf{G}_\text{CC}^\text{r}(E,x) \mathbf{\Gamma}_\text{R}(E,x) \mathbf{G}_\text{CC}^\text{a}(E,x)\right].\label{eq:tauEx_full_appendix}
\end{equation} 
Here $\mathbf{G}_\text{CC}^\text{r}$ denotes the retarded Green's function of the molecule, which is related to the advanced Green's function $\mathbf{G}_\text{CC}^\text{a}=(\mathbf{G}_\text{CC}^\text{r})^\dagger$ by Hermitian conjugation, and $\mathbf{\Gamma}_X=-2\mathrm{Im}(\mathbf{\Sigma}^\text{r}_{X})$ is the linewidth broadening matrix of electrode $X=\text{L},\text{R}$, which is derived from the retarded self-energy $\mathbf{\Sigma}^\text{r}_X$. Bold symbols here denote matrices.

To simplify the expressions, we make use of the wide-band approximation, yielding energy-independent self-energies and linewidth broadening matrices for $E$ in the vicinity of the Fermi energy $E_\text{F}$. We further assume that both electrodes interact only with terminal anchor atoms, which establish the covalent bonds at each side. If there is only a single atomic orbital relevant for transport at the Fermi energy at each side `l' and `r' of the junction in Fig.~\ref{fig:junction}, we finally obtain
\begin{equation}
    \tau(E,x) = \gamma_\text{L}\gamma_\text{R}\lvert G_\text{lr}^\text{r}(E,x)\rvert^2.\label{eq:tauEx_scalarG_appendix}
\end{equation} 
In the expression, $G_\text{lr}^\text{r}$ is the retarded Green's function of Eq.~\eqref{eq:G0r-spectral}, and $\gamma_\text{L}, \gamma_\text{R}$ parameterize electronic couplings of molecular anchoring orbitals to the electrodes. The dependencies of the transmission on energy $E$ and control parameter $x$ now result basically from the Green's function of the molecule. 

Only one transmission eigenchannel is present in
Eq.~\eqref{eq:tauEx_scalarG_appendix} due to the assumed coupling of the
molecule to the electrode by just one electronic orbital on the left and right
side. Similarly the  four-level model in the main text admits only one
transmission eigenchannel, and the total transmission thus equals the
transmission of the first eigenchannel, i.e.\ $\tau(E,x)=\tau_1(E,x)$. In
contrast, in quantum transport calculations based on density functional theory
(DFT), several transmission eigenchannels $i$ might contribute to
$\tau(E,x)=\sum_i \tau_i(E,x)$ in
Eq.~\eqref{eq:tauEx_full_appendix}.\cite{paulyClusterbasedDensityfunctionalApproach2008}
The zeros of a simplified four-level model, leading to suppressions of
$\tau_1(E,x)$, may then be detected in the DFT results by searching for points
in the $(E,x)$ plane, where transmissions of first and second eigenchannels
are similar, i.e.\ $\tau_1(E,x)\approx\tau_2(E,x)$. The presence of more than
one transmission eigenchannel will typically lead to an incomplete suppression of $\tau$ at the destructive quantum interference.

\section*{Author declarations}

\subsection*{Conflict of Interest}

The authors have no conflicts to disclose.

\section*{Data availability}

The data supporting the findings of this study are available within the article or can be obtained from the corresponding author upon reasonable request. 

\section*{References}
\bibliography{ClassifyingDQI}

\end{document}